\documentclass{article}
\usepackage[preprint]{spconf}
\usepackage{amsmath,hyperref}
\usepackage{fancyhdr}

\usepackage[dvipdfmx]{graphicx}

\title{SONAR: Self-Distilled Continual pre-training \\for Domain Adaptive Audio representation}

\name{Yizhou Zhang, Yuan Gao, Wangjin Zhou, Zicheng Yuan, Keisuke Imoto, Tatsuya Kawahara}

\address{
Graduate School of Informatics, Kyoto University, Japan \\
\texttt{yizhang@sap.ist.i.kyoto-u.ac.jp, keisuke.imoto@ieee.org}
}

\begin{document}
\ninept
\maketitle
\begin{abstract}
Self-supervised learning (SSL) on large-scale datasets like AudioSet has become the dominant paradigm for audio representation learning. While the continuous influx of new, unlabeled audio presents an opportunity to enrich these static representations, a naive approach is to retrain the model from scratch using all available data. However, this method is computationally prohibitive and discards the valuable knowledge embedded in the previously trained model weights. To address this inefficiency, we propose SONAR (\textbf{S}elf-distilled c\textbf{ON}tinual pre-training for domain adaptive \textbf{A}udio \textbf{R}epresentations), a continual pre-training framework built upon BEATs. SONAR effectively adapts to new domains while mitigating catastrophic forgetting by tackling three key challenges: implementing a joint sampling strategy for new and prior data, applying regularization to balance specificity and generality, and dynamically expanding the tokenizer codebook for novel acoustic patterns. Experiments across four distinct domains demonstrate that our method achieves both high adaptability and robust resistance to forgetting.
\end{abstract}

\vspace{-3pt}
\section{Introduction}
\vspace{-7pt}
Self-supervised learning (SSL) has emerged as a powerful paradigm for representation learning in the audio domain. Unlike supervised methods that require extensive human annotations, SSL leverages large-scale unlabeled data to learn general-purpose representations. This paradigm has driven remarkable progress in audio analysis, and has become the foundation for state-of-the-art audio models \cite{liu2022audio}. SSL depends critically on diverse and large-scale datasets. In practice, however, such datasets inevitably yield long tail distributions, for instance, AudioSet \cite{gemmeke2017audioset}, where many domains are present but insufficiently represented, leaving models unfamiliar with underrepresented domains. Compounding this issue, real-world data emerges as a continuous stream from heterogeneous sources, spanning vastly different acoustic domains like human speech, music, animal calls, and environmental sounds. Conventional solutions retrain models from scratch on all old and new data, yet this approach is computationally expensive and discards the valuable knowledge embedded in existing models \cite{bharadwaj2025openbeats}.

Continual pre-training offers a promising alternative, enabling models to incrementally adapt to new domains without discarding prior knowledge. Yet, applying audio SSL in a continual setting, where the goal is to learn from new data without access to all previous data, introduces three unique challenges. First, the audio domain is characterized by highly diverse and heterogeneous data streams, making it non-trivial to sample and organize training data effectively \cite{shi2024wavespecenc}. Second, successful domain adaptation hinges on achieving a delicate balance between specificity and generality. The model must integrate new domain-specific knowledge while preserving generalizable representations from prior data, a trade-off managed through regularization to prevent catastrophic forgetting \cite{kecontinual}. Third, the nature of raw audio poses a significant tokenization challenge. Unlike text, audio is a continuous signal that lacks a predefined vocabulary or clear segmentation and often contains multiple overlapping acoustic units. This inherent complexity means a fixed codebook is brittle when encountering new domains, as it cannot adequately represent novel acoustic patterns. Therefore, our framework necessitates a dynamically expandable codebook to ensure that learned representations remain both expressive and domain-relevant.

To address these challenges, we introduce SONAR, a continual self-supervised learning framework built upon the BEATs architecture. BEATs is a powerful self-supervised model that tokenizes raw audio into discrete semantic units and learns rich representations via a masked prediction task, akin to BERT in natural language processing \cite{chen2023beats}. We selected BEATs as our foundation for two key reasons. First, its unique combination of audio tokenization and a high masking-ratio self-distillation objective makes it an excellent model for general-purpose audio. More importantly for our work, this self-distillation structure provides a robust framework for stable model adaptation, making it highly suitable for the challenges of continual pre-training.

SONAR enhances this foundation through a multi-level framework that synergistically addresses the core challenges at the data, learning, and model levels. At the data level, we introduce Task-Relevant Stratified Sampling, which actively constructs a balanced training corpus by retrieving relevant examples from both a general knowledge pool and domain-specific data. At the learning level, we apply Dual-source Self-Distillation, a hierarchical regularization strategy that stabilizes both the low-level acoustic tokenizer and the high-level semantic encoder to combat catastrophic forgetting. Finally, at the model level, we implement an Online Clustered Codebook, which transforms the static tokenizer into a dynamic vocabulary that evolves to represent novel acoustic patterns by reinitializing underused codes.

We evaluate SONAR on four domain-specific cases spanning speech, music, bioacoustics, and environmental sound. Experimental results demonstrate that our framework achieves high adaptability and strong resistance to forgetting.

\begin{figure*}[htb]
  \vspace{-5pt}
  \centering
  \includegraphics[width=0.84\textwidth]{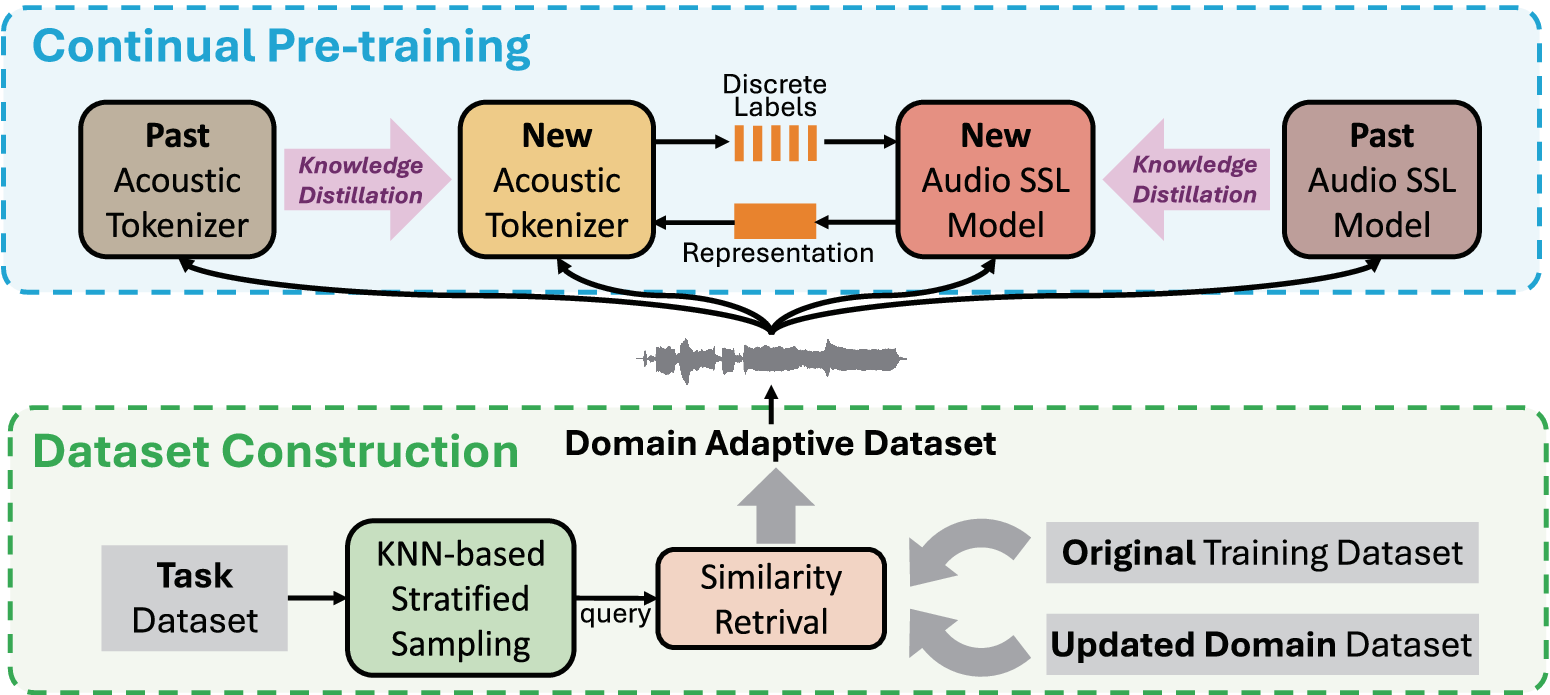}
  \vspace{-5pt}
  \caption{Overview of the SONAR framework for continual self-supervised audio representation learning, as described in Section 3. This framework integrates task relevant stratified sampling (Section 3.1), dual-source self-distillation (Section 3.2), and an online clustered codebook (Section 3.3) for dynamic adaptation to novel acoustic patterns. The approach enables efficient model adaptation across multiple domains while mitigating catastrophic forgetting.}
  \label{fig:mp}
\vspace{-4pt}
\end{figure*}

\vspace{-8pt}
\section{Related Work}
\vspace{-7pt}
\subsection{Self-Supervised Learning}
\vspace{-5pt}
Self-supervised learning (SSL) has become a cornerstone of audio representation learning, enabling models to leverage vast amounts of unlabeled data. Early SSL approaches in speech, such as wav2vec \cite{schneider2019wav2vec}, demonstrated the potential of contrastive prediction objectives. Building on this line, wav2vec 2.0 \cite{baevski2020wav2vec} and HuBERT \cite{hsu2021hubert} advanced the field by integrating masked prediction and discrete units, showing strong improvements across speech recognition and speaker tasks. By treating audio as spectrogram, influential SSL frameworks like AST \cite{gong2021ast}, and AudioMAE \cite{huang2022masked} have established that large-scale pre-training is highly effective for learning powerful, general-purpose representations. More recently, BEATs \cite{chen2023beats} introduced a self-distillation framework with a tokenizer and masked autoencoder, offering strong cross domain transferability. However, such models are typically trained in an offline setting on a fixed corpus, whereas real-world audio often arrives as a continuous stream from heterogeneous domains. To address this gap, our contribution is to develop a continual self-supervised framework that incrementally adapts BEATs to streaming audio while mitigating forgetting.

\vspace{-8pt}
\subsection{Continual pre-training}
\vspace{-4pt}
While continual learning has been extensively studied in natural language and vision via exemplar replay~\cite{rebuffi2017icarl}, regularization~\cite{kirkpatrick2017overcoming, li2016learning}, and modular architectures~\cite{houlsby2019parameter}, recent efforts have extended these ideas to self-supervised continual pre-training. For example, large language models are incrementally adapted to domain-specific corpora while retaining general knowledge~\cite{reed2022self, parmar2024reuse}.

In the broader audio domain, continual learning has also been applied to analysis tasks such as acoustic scene classification and audio tagging. These studies often adapt established methods like experience replay to handle scenarios with new sound classes or changing recording conditions~\cite{mulimani2025domain}. However, continual \textit{pre-training} in audio has so far focused mainly on speech, typically building upon wav2vec 2.0~\cite{baevski2020wav2vec}. Existing methods fall into three broad categories. Data-driven approaches seek to mitigate forgetting by controlling input distribution over time, such as using episodic memory buffers to replay past samples~\cite{yang2022online}, or scheduling training data across domains to reflect low-resource adaptation scenarios~\cite{getman2024continued}. Regularization-based methods, including CTRL and MCTRL~\cite{lee2022ctrl, choi2023m}, stabilize learning by integrating elastic weight consolidation (EWC), knowledge distillation with momentum-based teacher models. Architecture-centric designs introduce modular components to isolate task or domain-specific knowledge, such as inserting lightweight adapters into HuBERT \cite{hsu2021hubert} to enable parameter-efficient transfer across evolving automatic speech recognition (ASR) tasks~\cite{shi2024dualpath, shi2024investigation}. In contrast to these prior works, our setting considers continual pre-training for general, non-speech audio, encompassing heterogeneous domains where data distributions are more diverse and lacking linguistic supervision.

\vspace{-7pt}
\section{Method}
\vspace{0pt}
SONAR is a self-distilled continual pre-training framework built upon BEATs, a powerful and widely-used SSL-based audio representation model~\cite{chen2023beats}. SONAR is a principled framework where each component is specifically designed to solve a core challenge of continual audio pre-training. To address the challenge of data organization in a continuous stream, SONAR employs task-relevant stratified sampling at the data level to construct a balanced and informative training corpus. To tackle the fundamental problem of catastrophic forgetting, it applies dual-source self-distillation regularization at the learning level to stabilize both the tokenizer and the encoder. Finally, to overcome the limitation of a fixed vocabulary, SONAR implements an online clustered codebook at the model level, allowing the model's vocabulary to dynamically evolve and capture novel acoustic patterns. These three modules are synergistically integrated: the sampling strategy provides a balanced data stream, the regularization controls the learning process, and the dynamic codebook ensures the model has the capacity to represent the new information. This integrated design enables progressive adaptation while ensuring stability, as summarized in Figure~\ref{fig:mp}.
\vspace{-8pt}
\subsection{Task Relevant Stratified Sampling}
\vspace{-2pt}
To construct the domain-adaptive dataset for continual pre-training, we employ a retrieval-based pipeline in a representation space induced by the old tokenizer. For each audio sample, we extract a fixed-dimensional sample-level embedding using the frozen tokenizer, which is kept unchanged across stages to ensure representations alignment. Based on the embeddings of the task-domain data, we apply KMeans clustering to obtain a coarse partition of the task embedding space, and uniformly sample a small number of instances from each cluster as queries, encouraging coverage across diverse regions of the task distribution.

For each selected query, we perform K nearest-neighbor retrieval from the domain dataset using cosine similarity between normalized embeddings. The retrieved samples are merged with the query instances and deduplicated to form the domain-adaptive dataset used in the next continual pre-training stage. This retrieval-based augmentation reinforces domain-relevant patterns while maintaining consistent token semantics throughout continual pre-training.
\vspace{-16pt}
\subsection{Dual-source Self-Distillation Regularization}

To achieve continual adaptation while preserving prior knowledge, we build upon the original self-distillation in BEATs and introduce two additional constraint-based distillation objectives. Specifically, we add a tokenizer-level constraint that stabilizes discrete token generation and a model-level constraint that preserves high-level semantic representations. This design mitigates catastrophic forgetting during continual pre-training.

\vspace{-10pt}
\subsubsection{Tokenizer-Level Regularization}
\vspace{-4pt}
At the tokenizer level, the goal is to preserve the stability of discrete tokens while adapting to new data. The loss is decomposed into three components.

First, the alignment loss is defined as the cosine similarity between the output of the tokenizer estimator and the output of the teacher model:
\vspace{-2pt}
\begin{equation}
\mathcal{L}_1 = \sum_{t=1}^T \cos(o_t, \hat{o}_t),
\vspace{0pt}
\end{equation}

\noindent where $o_t$ is the output of the tokenizer estimator and $\hat{o}_t$ is the corresponding teacher feature, encouraging alignment with teacher representations.

Second, the vector quantization loss is defined as mean squared error (MSE) with a straight-through gradient mechanism:
\vspace{-7pt}
\begin{equation}
\mathcal{L}_2 = \sum_{t=1}^T \Big(
   -\|\ell_2(e_t) - \operatorname{sg}[\ell_2(v_{\hat{z}_t})]\|_2^2
   -\|\operatorname{sg}[\ell_2(e_t)] - \ell_2(v_{\hat{z}_t})\|_2^2
\Big),
\label{eq:vq_loss}
\vspace{2pt}
\end{equation}

\noindent where $e_t$ is the tokenizer's encoder output at time step $t$, $v_{\hat{z}_t}$ is the selected codebook vector, $\ell_2(\cdot)$ denotes $L_2$ normalization, and $\operatorname{sg}[\cdot]$ is the stop-gradient operation. This term enforces reconstruction through the codebook and stabilizes quantization.

Finally, the regularization loss is given by:
\vspace{-8pt}
\begin{equation}
\mathcal{L}_3 = \lambda_{\text{reg}} \sum_{t=1}^T 
      \|\ell_2(e_t) - \ell_2(\bar{e}_t)\|_2^2,
\vspace{-5pt}
\end{equation}

\noindent where $\bar{e}_t$ is the frozen encoder output from the previous tokenizer and $\lambda_{\text{reg}}$ controls the regularization strength, constraining the current encoder to remain close to its historical representation.

The overall tokenizer loss is then:
\vspace{-5pt}
\begin{equation}
\mathcal{L}_{\text{TOK}} = \mathcal{L}_1 + \mathcal{L}_2 + \mathcal{L}_{3}.
\vspace{-12pt}
\end{equation}

\vspace{-5pt}
\subsubsection{Model-Level Distillation}
\vspace{-5pt}
To stabilize semantic features in the BEATs encoder, we extend the Masked Audio Modeling (MAM) loss with a distillation constraint:
\vspace{-4pt}
\begin{align}
\mathcal{L}_{\text{MAM}} =
- \sum_{t \in \mathcal{M}} \log p(\hat{z}_t \mid \mathbf{X}_{\mathcal{M}^c})
+ \mu_{\text{reg}} \sum_{t=1}^T  \|\ell_2(r_t) - \ell_2(\bar{r}_t)\|_2^2 . \nonumber\\[-6pt]
\end{align}
\vspace{-18pt}

The first term is the standard cross-entropy loss for masked prediction, where the model predicts the token index $\hat{z}_t$ for masked steps $\mathcal{M}$ given unmasked context $\mathbf{X}_{\mathcal{M}^c}$. The second term is a feature-level distillation regularizer, penalizing the squared $L_2$ distance between normalized encoder representation of the current model $r_t$ and the frozen model $\bar{r}_t$. The hyperparameter $\mu_{\text{reg}}$ controls the strength of this constraint.
\vspace{-12pt}
\subsection{Online Clustered Codebook}
\vspace{-5pt}
We adopt an online clustered codebook update strategy, inspired by CVQ-VAE~\cite{Zheng_2023_CVQ}, to adapt to domain-specific acoustic variations while preventing codebook collapse. This design complements the regularization in Section~3.2 by ensuring stable and discriminative discrete tokens through dynamic updates.

The vector quantization loss is extended to include commitment and embedding update terms, as defined in Section~3.2 (Eq.~\ref{eq:vq_loss}). To maintain codebook utilization, we track assignment counts using an exponential moving average:
\vspace{-9pt}
\begin{equation}
N_k^{(t)} = \gamma N_k^{(t-1)} + (1 - \gamma)\cdot \frac{n_k^{(t)}}{BHW},
\vspace{-7pt}
\end{equation}
where $n_k^{(t)}$ is the number of tokens assigned to $v_k$ in batch $t$, and $BHW$ denotes the total number of tokens in the batch, with $B$ as the batch size and $H, W$ as the spatial dimensions of the feature map. Underused codes are softly reinitialized toward the current feature centroid $\tilde{e}_k^{(t)}$:
\vspace{-10pt}
\begin{equation}
v_k^{(t)} = (1 - \alpha_k^{(t)})\, v_k^{(t-1)} + \alpha_k^{(t)}\, \tilde{e}_k^{(t)},
\vspace{-4pt}
\end{equation}
where $\alpha_k^{(t)}$ is a reinitialization factor based on $N_k^{(t)}$.

To further enhance code discrimination, we introduce a contrastive loss that encourages assigned features to stay closer to their matched code $v_k$ than to negatives:
\vspace{-6pt}
\begin{equation}
\mathcal{L}_{\text{contra}} = 
-\log \frac{\exp\big(\operatorname{sim}(v_k, e_t^+)/\tau\big)}
{\sum_j \exp\big(\operatorname{sim}(v_k, e_j^-)/\tau\big)},
\vspace{-5pt}
\end{equation}
where $e_t^+$ is a positive feature assigned to $v_k$, $e_j^-$ are negatives from other codes, $\operatorname{sim}(\cdot,\cdot)$ denotes cosine similarity, and $\tau$ is the temperature.

The final codebook loss is:
\vspace{-6pt}
\begin{equation}
\mathcal{L}_{\text{codebook}} = \mathcal{L}_{\text{TOK}} + \lambda_{\text{contra}} \mathcal{L}_{\text{contra}} .
\vspace{-2pt}
\end{equation}
\vspace{-20pt}
\section{Experiments}
\vspace{-10pt}
\subsection{Datasets}
\vspace{-6pt}
In our experiments, the objective is to evaluate continual pre-training in diverse settings. To this end, we adapt across four domains using unlabeled audio only.
We construct EMO by aggregating speech-emotion corpora from CREMA-D~\cite{cao2014crema}, 
MELD~\cite{poria2019meld}, and CMU-MOSEI~\cite{zadeh2018multimodal} (about 40k utterances; labels discarded). 
FMA Large provides 106k music tracks spanning 161 genres~\cite{Defferrard2016FMAAD} (hereafter FMA). 
iNaturalist Sounds contains about 230k bioacoustic recordings~\cite{chasmai2024inaturalist} (hereafter iNat). 
FreeSound, taken from the WavCaps subset~\cite{mei2023wavcaps}, consists of about 262k environmental clips (hereafter FSD). 
Together these corpora cover speech, music, bioacoustics, and environmental sound, aligning with our target continual adaptation scenario.

For downstream evaluation, we use four representative datasets: 
IEMOCAP for speech emotion recognition~\cite{busso2008iemocap}, 
GTZAN for music genre classification~\cite{tzanetakis2002musical}, 
CBI for bioacoustics~\cite{hagiwara2023beans}, 
and TAU Urban Acoustic Scenes for environmental sound~\cite{heittola2020acoustic}. 
We use official splits when available\cite{turian2022hear}; otherwise we adopt a 3:1:1 train/validation/test division.
\begin{table}[t]
\vspace{-7pt}
\small
\centering
\caption{Comparison of methods on downstream tasks.}
\label{tab:acc_comp}
\begin{tabular*}{0.46\textwidth}{@{\extracolsep{\fill}}lcccc@{}}
\hline
\ \\[-9pt]
\textbf{Method} & \textbf{IEM} & \textbf{GTZ} & \textbf{CBI} & \textbf{TAU} \\
\hline
\multicolumn{5}{l}{\textit{Freezing}} \\
BEATs & 66.3 & 86.0 & 43.5 & 70.3 \\
DCPT & 62.8 & 76.0 & 11.9 & 64.4 \\
\textbf{SONAR (proposed)} & \textbf{69.0} & \textbf{88.0} & \textbf{44.2} & \textbf{70.4} \\
\ \ \ - Clustered Codebook & 68.0 & 87.5 & 44.2 & 70.4 \\
\ \ \ - Stratified Sampling & 68.7 & 86.9 & 43.1 & 69.8 \\
\ \ \ - Both Above & 67.8 & 86.5 & 42.9 & 68.3 \\
\hline
\multicolumn{5}{l}{\textit{Fine-tuning}} \\
BEATs & 68.4 & 82.0 & 64.7 & 78.6 \\
DCPT & 67.7 & 77.5 & 46.5 & 69.4 \\
\textbf{SONAR (proposed)} & \textbf{70.6} & \textbf{85.5} & \textbf{65.6} & 78.9 \\
\ \ \ - Clustered Codebook & 69.7 & 84.5 & 64.3 & \textbf{79.2} \\
\ \ \ - Stratified Sampling & 70.0 & 83.0 & 65.2 & 78.6 \\
\ \ \ - Both Above & 69.5 & 82.5 & 63.8 & 78.9 \\
\hline
\end{tabular*}
\vspace{-8pt}
\end{table}
\vspace{-9pt}
\subsection{Experimental Setup}
\vspace{-5pt}
For our experiments, we first construct domain-adaptive datasets for EMO, FMA, iNat, and FSD using task-relevant stratified sampling, yielding approximately 30k–50k audio segments per domain. For each domain, we separately adapt the BEATs iter3+ model on its respective dataset. All training experiments are conducted on NVIDIA RTX6000 Ada GPUs, with continual pre-training for each adaptation running for 10 epochs using the Adam optimizer and a learning rate of $1 \times 10^{-4}$, a regularization weight of $\lambda_{\text{reg}} = 1 \times 10^6$. Our clustered codebook update mechanism is configured with a contrastive loss weight of $\lambda_{\text{contra}} = 10$, an exponential moving average decay of $\gamma = 0.9$ for code usage tracking, and a contrastive temperature of $\tau = 0.3$. All hyperparameters were selected based on preliminary validation and are held fixed across all domains. For downstream evaluation, a linear classification head is attached to the adapted model. The adapted model is either frozen or fine-tuned, depending on the configuration, as shown in Table \ref{tab:acc_comp}.

\begin{table}[t]
\vspace{-7pt}
\centering
\caption{Knowledge retention on AudioSet.}
\label{tab:forgetting_eval}
\begin{tabular*}{0.46\textwidth}{@{\extracolsep{\fill}}lcc@{}}
\hline
\ \\[-9pt]
\textbf{Method} & \textbf{mAP} & \textbf{FR} \\
\hline
\multicolumn{1}{l}{\textit{Baseline}} \\
BEATs & 34.8 & 0.0 \\
\hline
\multicolumn{3}{l}{\textit{After EMO}} \\
DCPT & 13.7 & 60.6 \\
\textbf{SONAR (proposed)} & \textbf{34.9} & \textbf{-0.3} \\
\ \ \ - Clustered Codebook & 34.6 & 0.6 \\
\ \ \ - Stratified Sampling & 34.7 & 0.3 \\
\ \ \ - Both Above & 34.3 & 1.4 \\
\hline
\multicolumn{3}{l}{\textit{After FMA}} \\
DCPT & 14.7 & 57.76 \\
\textbf{SONAR (proposed)} & 34.7 & 0.3 \\
\ \ \ - Clustered Codebook & 34.2 & 1.7 \\
\ \ \ - Stratified Sampling & 34.4 & 1.2 \\
\ \ \ - Both Above & 33.8 & 2.9 \\
\hline
\multicolumn{3}{l}{\textit{After iNaturalist}} \\
DCPT & 12.5 & 73.5 \\
\textbf{SONAR (proposed)} & 34.5 & 4.2 \\
\ \ \ - Clustered Codebook & 33.8 & 6.1 \\
\ \ \ - Stratified Sampling & 34.2 & 5.0 \\
\ \ \ - Both Above & 33.6 & 7.0 \\
\hline
\multicolumn{3}{l}{\textit{After FreeSound}} \\
DCPT & 13.6 & 60.9 \\
\textbf{SONAR (proposed)} & 34.7 & 0.3 \\
\ \ \ - Clustered Codebook & 34.3 & 1.4 \\
\ \ \ - Stratified Sampling & 34.4 & 1.2 \\
\ \ \ - Both Above & 33.9 & 2.6\\
\hline
\end{tabular*}
\vspace{-10pt}
\end{table}
\vspace{-10pt}
\subsection{Evaluation Metrics}
\vspace{-5pt}
We evaluate SONAR from two complementary perspectives: 
its ability to acquire new domain knowledge (plasticity) and its capacity to retain prior knowledge (stability). 

Plasticity is assessed by the F1 score on downstream classification tasks, 
which provides a balanced measure under class imbalance. 

Stability is assessed by mean Average Precision (mAP) on the original AudioSet pre-training task, 
together with the forgetting rate (FR). 
Formally, FR is computed as the relative drop in mAP with respect to the baseline model: 
\vspace{-4pt}
\[
\mathrm{FR} = \frac{\mathrm{mAP}_{\text{baseline}} - \mathrm{mAP}_{\text{current}}}{\mathrm{mAP}_{\text{baseline}}} \times 100\% .
\vspace{-4pt}
\]
Here, $\mathrm{mAP}_{\text{baseline}}$ denotes the performance of the original BEATs model on AudioSet before continual pre-training, 
and $\mathrm{mAP}_{\text{current}}$ is the corresponding performance after adaptation. 
Lower FR indicates stronger knowledge retention.
\vspace{-8pt}
\subsection{Results and Ablation Study}
\vspace{-5pt}
The proposed method consistently outperformed the baseline models, BEATs and Direct Continual pre-training (DCPT, which resumes training from the original BEATs checkpoint on new domain data using the same pre-training objectives), across four distinct audio domains, as shown in Table~\ref{tab:acc_comp}. It achieved F1 scores of 70.6\% on speech emotion recognition with IEMOCAP, 88.0\% on music genre classification with GTZAN, 65.6\% on bioacoustic call identification with CBI, and 78.9\% on environmental sound classification with TAU, which highlights SONAR's strong knowledge transfer capabilities.

Beyond the improvement on new tasks, SONAR's ability to overcome catastrophic forgetting was tested on the AudioSet dataset, which is an audio tagging task where each clip may be assigned multiple semantic labels. As summarized in Table~\ref{tab:forgetting_eval}, it successfully retained prior knowledge, maintaining a mean Average Precision (mAP) between 34.5 and 34.9, with a forgetting rate close to zero. In stark contrast, the DCPT model suffered from severe catastrophic forgetting, with its mAP dropping to 12.5, a forgetting rate of over 70\%. This result demonstrates that SONAR can effectively integrate new knowledge while preserving existing representations.

The model's robust performance is further explained by an ablation study, with results showed in Table \ref{tab:acc_comp} and \ref{tab:forgetting_eval}. These tables illustrate how performance metrics show a clear, monotonic improvement trend in most cases as SONAR's key modules are progressively integrated. This step-by-step enhancement confirms the contribution of its core components, while also revealing that the importance of each module can vary depending on the specific dataset: dual-source self-distillation first stabilizes the model's features; task-relevant stratified sampling is then added to enhance generalization, which is particularly effective on datasets with significant distribution shifts; finally, online clustered codebook updates reinforce representational continuity. The generally consistent performance gains shown in the tables validate that this combination allows the full SONAR framework to achieve the best trade-off between adaptability and stability.
\vspace{-18pt}
\section{Conclusion}
\vspace{-8pt}
In this paper, we introduced SONAR, a continual pre-training framework designed to address the dual challenge of enhancing model adaptability to new domains while preventing catastrophic forgetting. By integrating task relevant stratified sampling, dual-source self-distillation, and an online cluster codebook, our method enables models to effectively learn novel acoustic patterns from unlabeled data without compromising previously acquired knowledge. Extensive experiments across four distinct domains demonstrated that SONAR achieves a superior balance, consistently outperforming baselines in both downstream task accuracy and resistance to forgetting, as measured by mAP.


\vspace{-11pt}
\section{Acknowledgments}
\vspace{-9pt}
This work was partially supported by JSPS KAKENHI Grant Numbers 22H03639 and 23K16908.

\vspace{-6pt}
\bibliographystyle{IEEEbib}
\bibliography{strings,refs}
\end{document}